\documentclass[preprintnumbers,nofootinbib,showpacs,twocolumn,aps,prl]{revtex4}

\usepackage{amsmath}
\usepackage{amssymb}
\usepackage{amsfonts}
\usepackage{graphicx}
\usepackage{epsfig}
\usepackage{subfigure}
\usepackage{color}

\begin{document}
\title{Algebraic Singularity Method for Mass Measurements
with Missing Energy}
\author{Ian-Woo Kim}
\affiliation{
Department of Physics, University of Wisconsin, Madison, WI 53706, USA} 
\preprint{MADPH-09-1539}
\date{\today}
\begin{abstract}
We propose a novel generalized method for mass measurements based on phase space singularity structures that can be applied to any event topology with missing energy.   
Our method subsumes the well-known end-point and transverse mass methods and yields new techniques for studying ``missing particle" events, such as the double chain production of stable neutral particles
at the LHC.  
\end{abstract}
\pacs{11.80.Cr,12.60.-i}
\maketitle

\textit{\underline{Introduction}}. 
At the Large Hadron Collider (LHC), physics beyond the Standard Model (SM) may reveal itself in signals with large missing transverse energy.  The reason is that many 
TeV scale models require new symmetries that distinguish the new particles from the SM fields,  which can result in stable neutral particles that can be attractive 
dark matter candidates.  One example is the neutralino lightest supersymmetric particle (LSP) in the minimal supersymmetric Standard Model (MSSM).  
The cascade decays of superpartners result in final state LSP's that are ``missing particles" that 
escape detection.

Measuring the masses of these missing particles is of great importance but is challenging.   The  full 
S-matrix can be used to determine the masses through the event distribution profile, but this is  
 highly model-dependent.  Mass measurement techniques thus rely on kinematic analysis, which does not allow for global fits using the event profile, but provide useful  information via the phase space structure defined by the kinematic constraints.

Event topologies with missing energy may or may not be directly reconstructable.  
For reconstructable events, the number of constraints equals or exceeds the number of invisible particle momentum components.
By reconstructing the momenta, the likelihood of a given test mass parameter for each individual event can be obtained \cite{reconstruction}.  
Though powerful, this method can be used only for exclusive processes and there are typically many combinatoric factors due to the large number of particles.

For non-reconstructable processes, it is not possible to obtain a likelihood contribution from individual events or fit the event profile globally using only kinematics. 
Mass measurement techniques include using the end-points of kinematic variables such as the invariant mass distribution \cite{invariantmass},   
or kinematic cusps \cite{Han:2009ss}.  Recently, there has been an emphasis on ``implicit'' variables that depend on trial masses~\cite{implicit}, such as 
the end-point of the $m_{T2}$ distribution, which has a kink when the trial masses equal the true masses~\cite{kink}.  Momentum reconstruction is also possible for events near the $m_{T2}$ end-point 
\cite{Cho:2008tj}, and there are attempts to understand the $m_{T2}$ kink based on end-points~\cite{boundary}.

It is not an accident that these methods use end-points, cusps, and kinks, which are {\em singularities} in the observable phase space. 
In this letter, we develop the general theory of kinematic singularities and provide a systematic method for obtaining new implicit optimized variables  that best exploit the singularity structure.  We begin with a systematic analysis of phase space singularities.  We then construct kinematic variables that we call  {\em singularity coordinates} and apply the method to cascade decays and double missing particle chains to show how previously studied examples are unified within our approach. The mathematical details and additional examples will 
be presented in a future publication~\cite{preparation}.

\begin{figure}
\includegraphics[width=7cm]{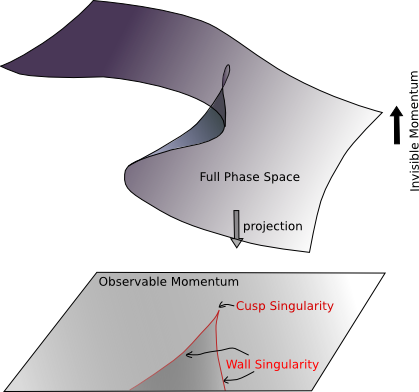}
\caption{A schematic diagram describing the relation between the full phase space and the projected observable phase space.}
\label{phasespace}
\end{figure}

\textit{\underline{Kinematic Singularities}}.
A singularity is a point where the local tangent space cannot be defined as
a plane, or has a different dimension than the tangent spaces at 
non-singular points.  
The full phase space does not exhibit singularities (we do not consider cases of singularities due to soft or collinear massless particles).  However, if only a subset of the momenta are measured, the relevant quantity is the projection of the full phase space on to the observable phase space of the measured momenta. Each observable phase space point can correspond to multiple configurations in the full phase space, as in Fig.~\ref{phasespace}. The {\em multiplicity} volume 
changes abruptly upon crossing a boundary where phase space folding occurs. The observable phase space then has a non-smooth density, reminiscent of caustics in optics. 

Given the visible momenta $q_j (j = 1,\dots,m)$ and the invisible momenta $x_k ( k = 1,\dots,n )$, 
the full phase space is embedded in $(n+m)$-dimensional Euclidean 
space as the solution space of the $N$ constraints:
\begin{eqnarray}
g_i (x,q) = 0,\qquad (i = 1,\dots,N),
\end{eqnarray}
where the $g_i$ are coupled polynomial equations that are at most of quadratic degree. In mathematics terminology, such a space is called an {\em affine variety}.

At the singularity, at least one direction of the tangent plane in the full phase space is aligned vertically along the invisible momentum direction.  The invisible momentum components of the normal vector that defines this tangent plane are given by the row vectors of the ``restricted'' Jacobian matrix $(\partial g_i / \partial x_k)$.  The vertical alignment of the tangent plane implies that at a singularity, 
the restricted Jacobian matrix 
must have a {\em reduced rank}: 
\begin{eqnarray}
{\rm Rank} \left(\frac{\partial g_i}{\partial x_k}\right)_{\rm sing}
< {\rm Rank} \left(\frac{\partial g_i}{\partial x_k}\right)_{\rm reg}.
\end{eqnarray}
The amount that the rank is reduced is the {\em degree} of the singularity; 
a wall (cusp) has degree one (two).

Finding the reduced rank condition of an arbitrary matrix is not an easy problem. 
However, 
for the special case of an affine variety, 
the given set of polynomial equations can be substituted by another set with the same solution space. The set of all such equivalent equations is
called an {\em ideal}, which is generated by a finite set of polynomials known as a {\em basis}.   We focus here on the  {\em  Gr\"obner basis}, in which variables are eliminated sequentially as follows: 
\begin{eqnarray}
g_1 (x_1, x_2, x_3, \dots, x_n) &=& 0, \nonumber \\
g_2 (x_2,x_3, \dots, x_n) &=& 0, \nonumber \\
&\vdots&\nonumber\\
g_N (x_N,x_{N+1}, \dots, x_n) &=& 0.
\end{eqnarray}
The algorithm for finding the Gr\"obner basis for a general coupled polynomial system is known~\cite{buchberger}. For the processes of interest, it is tractable to obtain it analytically.  

The reduced rank condition 
implies that one or more row vectors of the restricted Jacobian are linearly dependent.  In the Gr\"obner basis, the restricted Jacobian is of upper triangular form.  Therefore, a necessary but not sufficient condition for linear dependency is that one of the diagonal components vanishes, resulting in an analytic condition for the singularity position.  
\vskip 0.3cm

\textit{\underline{Singularity Coordinates}}.~
The next step is to construct an optimized one-dimensional variable that we call the  {\em singularity coordinate.} This is an implicit variable because the location of the singularity is defined by the reduced rank condition of the restricted Jacobian matrix, which is an implicit 
function of the mass parameters. 
The singularity coordinate must satisfy the following criteria: (i)
it must be zero at the singularity, (ii) its direction must be perpendicular to 
the singularity hypersurface in observable phase space, and (iii)
it must be normalized such that every event can give 
the same significance.
 
To see this, we note that the reduced rank condition implies that one linear combination $\sum_i c_i (\partial g_i / \partial x_k)$ becomes a null vector at the singularity point.  
The perpendicular direction is determined by
$\left(\vec{v}\right)_j = \sum_i c_i (\partial g_i / \partial q_j)$,
(recall $q_j$ are the visible momenta).
The singularity coordinate in this direction is maximally efficient for revealing the singularity structure.
To assign an unambiguous value to each event,  the singularity coordinate is scaled so that events with the same invisible phase space volume around the nearest singularity have the same value.
This requires a knowledge of the local phase space properties around the singularity at quadratic order. 

\begin{figure}
\centering
\includegraphics[width=7cm]{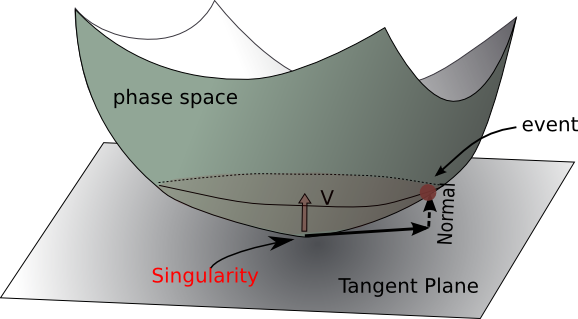}
\caption{The scaling behavior near a singularity.}
\label{singscale}
\end{figure}

As shown in Fig.~\ref{singscale},  a local orthonormal coordinate system around a given reference point can be split into tangent directions $t_r, (r=1,\dots n+m-N)$ 
and normal directions $n_s, (s=1, \dots, N)$. A general phase space point near
this reference point is labeled by the tangent coordinate.  The normal 
coordinate is determined by a quadratic function of the tangent coordinate (the {\em second fundamental form}): $n_s \equiv I\!I^s ( t^r ) = M^s_{r r'} t^r t^{r'}$, where  
\begin{eqnarray}
I\!I^s ( t_r )&=&  - \left( 
\frac{\partial g_p}{\partial n_s} \right)^{-1} 
 \frac{\partial^2 g_p}{\partial t_r \partial t_{r'}} t^r t^{r'}. 
\label{secondfundform}
\end{eqnarray}
We define $\tilde{\Sigma} \equiv \vec{v} \cdot I\!I (t_r) \equiv M_{rr'} t_r t_{r'}$. 
To find the appropriate scale factor, we need to obtain the phase space volume in the tangent directions that correspond to invisible momenta.  The phase space in the invisible momentum 
directions in the diagonalized basis $\tilde t_r$ is given by 
$a_1 \tilde{t}_1^2 
+ \dots + a_M \tilde{t}_M^2 = \tilde\Sigma$, 
where $M$ is the number of invisible tangent directions and the eigenvalues $a_r$ determine the shape of the invisible phase space around the singularity. 
For 
positive eigenvalues, 
the ellipsoid-shaped phase space volume scales as
$({\rm Vol}) \propto \left({a_1 a_2 \dots a_M}\right)^{-1/2} 
\tilde{\Sigma}^{M/2}$.
The 
singularity coordinate $\Sigma$ that satisfies all three criteria is thus given by
\begin{eqnarray}
\Sigma \equiv \left({a_1 \dots a_M}\right)^{-1/M} \tilde{\Sigma}. 
\end{eqnarray}
$\Sigma$ is an implicit kinematic variable, since the location of the zero, the normal direction
$\vec{v}$ and the scale factor can be defined 
only when mass parameters are given.
\begin{figure}
\includegraphics[width=5cm]{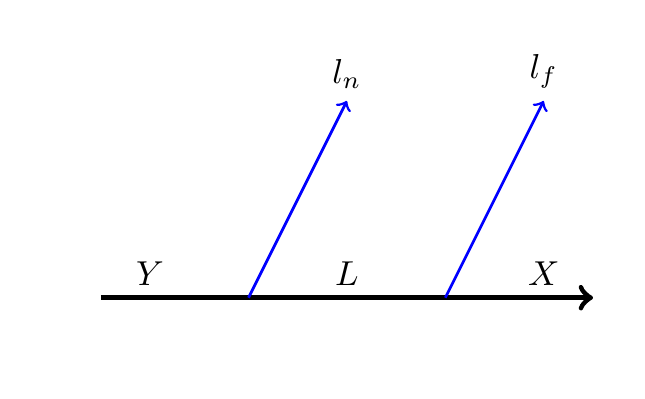}
\caption{The event topology of a simple cascade decay. }
\label{simplecascade}
\end{figure}

\textit{\underline{Simple Cascade Decay}}.~
Our first example is the simple cascade decay process shown in Fig.~\ref{simplecascade} (e.g. neutralino decay in the MSSM, with $Y=\tilde{\chi}_2^0$, $L=\tilde{l}$, and $X=\tilde{\chi}_1^0$). The on-shell equations of this system are
\begin{eqnarray}
x^2=m_{X}^2,\;\;
(x+q_f)^2=m_{L}^2,\;\; 
(x+q_f+q_n)^2=m_{Y}^2. \label{singlechainsystem}
\end{eqnarray}
$m_{X}$, $m_{L}$, and $m_Y$ are trial masses, $x$ is the invisible particle momentum, and $q_{n,f}$ are the visible particle momenta. 
Taking the $z$-axis in the direction of the 3-momentum of $l_n$ in the center of mass (CM) frame of the visible particles $l_{n,f}$, we have  
$q^{\rm cm}_{n,f} = (E_{\rm cm} / 2  , 0 ,0, \pm E_{\rm cm}/2  )$,
where $E_{\rm cm}$ is the CM energy of $l_{n,f}$.  
Eq.~(\ref{singlechainsystem}) is then
\begin{eqnarray}
&&x_0^2-x_1^2-x_2^2-x_3^2=m_{X}^2,\nonumber \\
&&(E_{\rm cm}/2 +x_0)^2-x_1^{2}-x_2^{2}
-(E_{\rm cm}/2+x_3)^2=m_{L}^2,\nonumber \\
&&(E_{\rm cm}+x_0)^2-x^2_1-x^2_2-x_3^2=m_{Y}^2. \label{singlechainsystem2}
\end{eqnarray}
The Gr\"obner basis for this system 
 is particularly
simple. With the lexicographic ordering $x_0 \succ x_3 \succ  x_1 \succ x_2 $, 
\begin{eqnarray}
g_1 &=&  2 E_{\rm cm} x_0 + \left(E_{\rm cm}^2 + m_X^2 - m_Y^2\right), 
\nonumber \\
g_2 
&=&  2 E_{\rm cm} x_3 + E_{\rm cm}^2 + 2 m_L^2 - m_X^2 - m_Y^2, \nonumber\\
g_3 &=&  E_{\rm cm}^2 x_1^2 +  E_{\rm cm}^2 x_2^2 \nonumber\\
&&+\left(   E_{\rm cm}^2 m_L^2  
  - (m_Y^2 - m_L^2 ) (m_L^2 - m_X^2) \right). \label{grobnerbasis} 
\end{eqnarray}
The restricted Jacobian $(\partial g_i  / \partial x_j)$ is 
\begin{eqnarray}
\left(
\begin{array}{cccc}
2 E_{\rm cm} &  & & \\
 & 2 E_{\rm cm} & & \\
 &             & 2 E_{\rm cm}^2 x_1 & 2 E_{\rm cm}^2 x_2   
\end{array}
 \right).
\end{eqnarray}
The first two row vectors are zero only when there are soft singularities, which we do not consider here.  The condition that the third vector vanishes results in 
$x_1 = x_2 = 0$.   Physically, this means that the missing particle momentum is aligned in the
direction of the lepton momentum in the CM frame. Together with Eq.~(\ref{singlechainsystem2}), 
this results in the following condition at the singularity:
\begin{eqnarray}
E^2_{\rm cm} = \frac{(m_Y^2 - m_L^2 ) ( m_L^2 - m_X^2 )}{m_L^2}
\equiv \left(m^{{\rm (max)}}_{ll}\right)^2.
\end{eqnarray} 
This reproduces the well-known result for the edge
of the invariant mass $m_{ll} \equiv \sqrt{(q_n + q_f)^2}$.  The tangent directions here are given by $x_{1,2}$, and $\vec{v}$ is the $E_{cm}$ direction.  The 
scale factor can be read off from $g_3=0$. 
The singularity coordinate is given by the well-known quantity    
 $\Sigma \propto ((m^{\rm max}_{ll})^2 -E_{\rm cm}^2)/(m^{\rm max}_{ll})^2$.
Our method also shows that no other singular structures exist for this process.
\begin{figure}
\centering
\includegraphics[width=5cm]{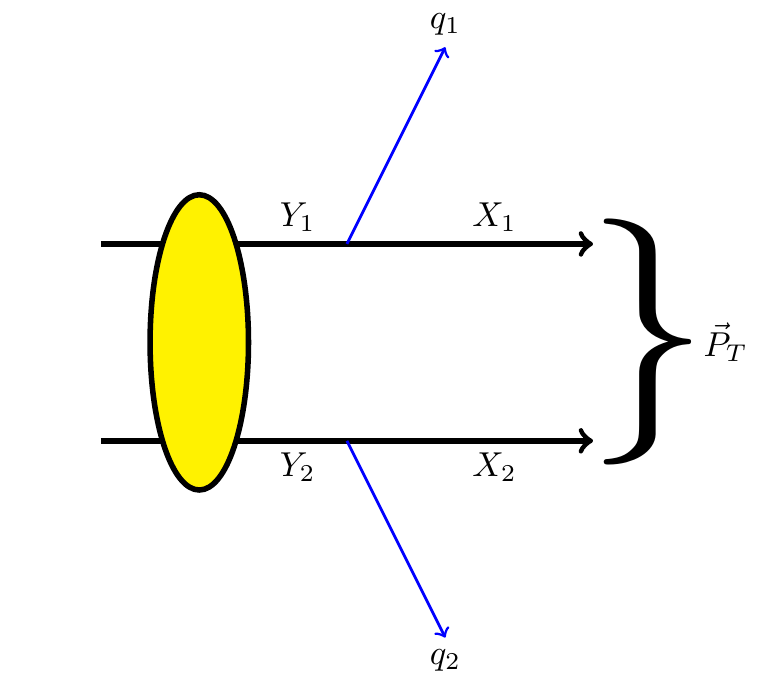}
\caption{The event topology of a double missing particle chain.}
\label{doublechain}
\end{figure}

\textit{\underline{Double Missing Particle Chain}}.~
For the double missing particle chain of Fig.~\ref{doublechain}, we have the relations
\begin{eqnarray}
x_1^2 = m_X^2, \quad x_2^2 = m_X^2, \quad
(x_1 + q_1)^2 = m_Y^2, \nonumber \\ (x_2+ q_2)^2 = m_Y^2, \quad 
\vec{x}_{1T} + \vec{x}_{2T} = \vec{p}_T,
\end{eqnarray}
where $x^\mu_i = (x_{i0}, x_{i1}, x_{i2}, x_{i3})$ denote the momenta of $X_i$,  
$q^\mu_{i} = (q_{i0}, q_{i1}, q_{i2}, q_{i3})$ are the visible
particle momenta for each chain, 
and  $\vec{p}_T = (p_{T1}, p_{T2})$ is the 
missing transverse momentum.  We assume $m_{X1} = m_{X2} = m_{X}$ and $m_{Y1}= m_{Y2} = m_{Y}$ (the generalization to an asymmetric chain is straightforward).
With 
$x_{10} \succ x_{13} \succ  x_{20} \succ  x_{21} \succ  x_{22} \succ  x_{23}  \succ x_{11} \succ  x_{12}$, the Gr\"obner basis takes the form
\begin{eqnarray}
g_1 &=& q_{10} x_{10} - q_{13} x_{13} - q_{11} x_{11} - q_{12} x_{12} 
- C_1 ,\\
g_2 &=& \left( q_{10}^2 - q_{13}^2 \right) x_{13}^2 
- 2 q_{11} q_{13} x_{13} x_{11} - 2 q_{12} q_{13} x_{13} x_{12}
\nonumber \\
&& - 2 C_1 q_{13} x_{13} + 
\left(q_{10}^2 - q_{11}^2 \right) x_{11}^2 
- 2 q_{11} q_{12} x_{11} x_{12} \nonumber \\
&&+ \left(q_{10}^2 - q_{12}^2 \right) x_{12}^2 
- 2 C_1 q_{11} x_{11} - 2 C_1 q_{12} x_{12} \nonumber \\
&& + \left( m_X^2 q_{10}^2 - C_1^2 \right),  \\ 
g_3 &=& q_{20} x_{20} - q_{23} x_{23} + q_{21} x_{11} 
 + q_{22} x_{12} - C_2, \\
g_4 &=& x_{21} + x_{11} -p_{T1}, \\
g_5 &=& x_{22} + x_{12} -p_{T2}, \\
g_6 &=& \left( q_{20}^2 - q_{23}^2 \right) x_{23}^2
+ 2 q_{21} q_{23} x_{23} x_{11} + 2 q_{22} q_{23} x_{23} x_{12}
\nonumber \\
&& - 2 C_2 q_{23} x_{23} 
+ \left(q_{20}^2 - q_{21}^2 \right) x_{11}^2
- 2 q_{21} q_{22} x_{11} x_{12} 
\nonumber \\
&&+ \left(q_{20}^2 - q_{22}^2 \right) x_{12}^2 \
-  \left( 2 p_{T1} q_{20}^2 - 2 C_2 q_{21} \right) x_{11} \\ 
&&-  \left( 2 p_{T2} q_{20}^2 - 2 C_2 q_{22} \right) x_{12} 
 + \left(\vec{p}_T^2  + m_X^2 \right) q_{20}^2 - C_2^2,  \nonumber
\end{eqnarray}
in which 
$C_1 = ( m_{Y}^2 - m_{X}^2 - q_1 \cdot q_1)/2$, $C_2 =( m_{Y}^2 - m_{X}^2 - q_2 \cdot q_2 +\vec{q}_{2T} \cdot \vec{p}_{T}$.
\begin{figure}[t]
\centering
\includegraphics[width=6cm,angle=90]{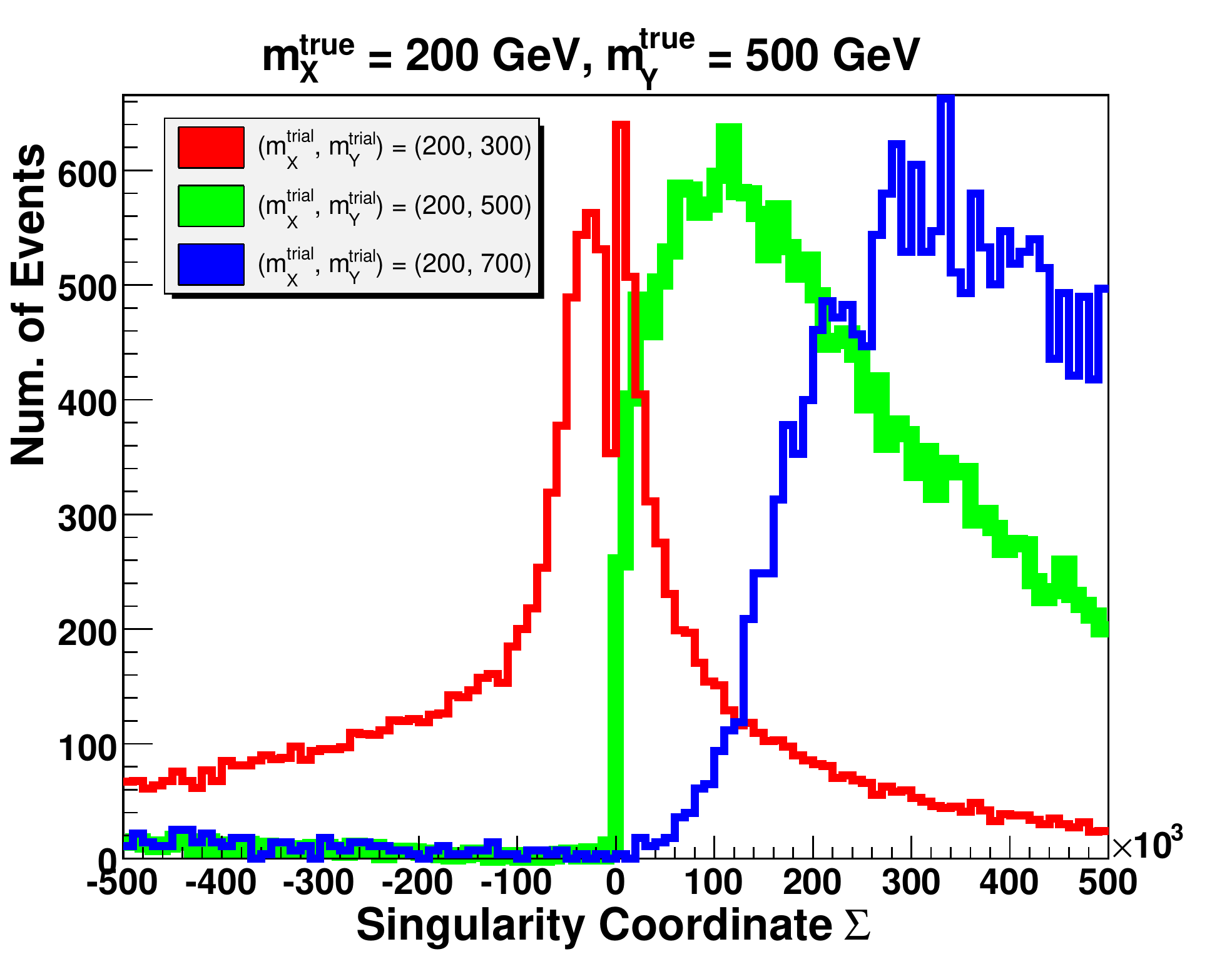}
\caption{The singularity coordinate distribution for several choices of trial masses 
with  $m_X = 200$  GeV, $m_Y = 500$  GeV. }
\label{result}
\end{figure}
The restricted Jacobian matrix $(\partial g_i / \partial x_j)$ 
with respect to the invisible momenta $x^\mu_{1,2}$ 
has the form 
\begin{center}
\begin{tabular}{c|cccccccc}
\hline
 & $x_{10}$ & $x_{13}$ & $x_{20}$ & $x_{21}$ &
$x_{22}$ & $x_{23}$ & $x_{11}$ & $x_{12}$ \\
\hline 
$g_1$ & $\Box$ & $\Box$ &   &   &   &   & $\Box$ & $\Box$ \\
$g_2$ &   & $\Box$ &   &   &   &   & $\Box$ & $\Box$ \\
$g_3$ &   &   & $\Box$ &   &   & $\Box$ & $\Box$ & $\Box$ \\
$g_4$ &   &   &   & 1 &   &   & 1 &   \\
$g_5$ &   &   &   &   & 1 &   &   & 1 \\
$g_6$ &   &   &   &   &   & $\Box$ & $\Box$ & $\Box$ \\
\hline
\end{tabular}
\end{center}
in which $\Box$ is a nonzero entry that depends on the visible and 
invisible momenta, $1$ is a constant (nonzero) term, 
and a blank space is a zero. 
The conditions $\partial g_1 / \partial x_{10} = q_{10}$ and
$\partial g_3 / \partial x_{20} = q_{20}$ correspond to soft singularities.  The restricted Jacobian $(\partial g_i / \partial x_j)$ has a reduced rank if
\begin{eqnarray}
\frac{\partial g_2}{\partial x_{13}}&=& 2 (q_{10}^2 - q_{13}^2) x_{13} 
 - 2  (C_1+\vec{q}_{1T} \cdot \vec{x}_{1T} )  q_{13}=0, \nonumber
\label{g2diag}  \\
\frac{\partial g_6}{\partial x_{23}}&=& 2(q_{20}^2 - q_{23}^2 ) x_{23}
 - 2 (C_2- \vec{q}_{2T} \cdot \vec{x}_{1T}) q_{23} =0, \nonumber  \\
&&\det 
\left( 
\begin{array}{cc}
\frac{\partial g_2}{\partial x_{11}} & 
\frac{\partial g_2}{\partial x_{12}} \\
\frac{\partial g_6}{\partial x_{11}} &
\frac{\partial g_6}{\partial x_{12}} 
\end{array}
\right) = 0. \label{reducedrank3}
\end{eqnarray}
Once the nearest singularity point is identified, we can determine the singularity coordinate.   
A numerical analysis for events with true masses $m_X = 200$  GeV and $m_Y = 500$  GeV is shown in Fig.~\ref{result}.   The singularity appears at $\Sigma=0$ only when the trial masses are equal to the true masses, providing a proof of concept.

Here we have neglected backgrounds and assumed that we have identified the event topology correctly.  
A complete treatment of the backgrounds must be done on a case-by-case basis and is beyond the scope of this paper.  However, as the singularity coordinates maximize any singular features at the true masses, they best discriminate the signal in the presence of backgrounds, which have smooth profiles at these points.  If the event topology was misidentified, singular features will not appear for any trial mass values, indicating that the hypothesis was incorrect.  Further details will be given in \cite{preparation}.

\textit{\underline{Conclusions}}.~
We have presented a systematic method for measuring missing 
particle masses based on phase space singularity structures
that is applicable to any non-reconstructable process.  
The approach reproduces well-known results, such as the end-point of the 
invariant mass distribution, and provides a means for finding 
the singularities for more general processes 
that can also be used for 
determining qualitative event properties. 
This method should provide invaluable tools in the search for new physics 
at the LHC.

\textit{\underline{Acknowledgments}}.~
We would like to thank A. Barr,  
W.~S.~Cho, K.~Choi, B.~Gripaios, T.~Han, C.~B.~Park, M.~Peskin,
Y.~Rao, J.~Song, L.~T.~Wang, and especially  L.~Everett for conversations and comments on the manuscript.   This work is supported by the U.S.
Department of Energy under grant No. DE-FG02-95ER40896.

\end{document}